\documentclass[11pt,dvips]{article}
\usepackage{epsfig,times} 
\makeatletter
\renewcommand{\section}{\@startsection{section}{1}{0in}
	{0.4\baselineskip}{0.1\baselineskip}{\Large\bf}}
\renewcommand{\subsection}{\@startsection{subsection}{2}{0in}
	{0.25\baselineskip}{-\baselineskip}{\large\bf}}
\renewcommand{\subsubsection}{\@startsection{subsubsection}{3}{0in}
	{0.1\baselineskip}{-\baselineskip}{\normalsize\bf}}
\makeatother
\begin{document}
\thispagestyle{myheadings}
\markright{OG.2.3.13}
\begin{center}
{\LARGE \bf Study of Gamma-Ray Bursts of energy $E>$10 GeV 

with the ARGO-YBJ detector}
\end{center}

\begin{center}

{\bf ARGO-YBJ Collaboration}, presented by S.Vernetto$^{1}$


{\it $^1$Istituto di Cosmo-Geofisica del CNR and INFN, Torino, Italy}

\vspace{1.cm}

{\large \bf Abstract\\}
\end{center}

The study of high energy gamma-ray bursts
can be performed by large area air shower arrays
operating at very high mountain altitudes.
ARGO-YBJ is a detector optimized to observe small size air showers,
to be constructed at the Yangbajing Laboratory (Tibet, China) at an
altitude of 4300 m.
One of the goals of the experiment is the study
of gamma-ray bursts of energies $E >$ 10 GeV.
This can be achieved using the "single particle" technique, more
profitable in the energy region $E<50$ GeV, and the "low multiplicity"
technique, suitable to observe GRBs at higher energies.
The sensitiviy of ARGO-YBJ allows the detection of GRBs 
with energy fluences in the range 1$\div$100 GeV as low as 
$F \sim 10^{-6} \div  
10^{-5}$ erg cm$^{-2}$, depending on the spectral characteristics
of the burst.

\vspace{-0.5ex}
%
\vspace{1ex}

\section{Introduction}

The observation of GeV photons 
by EGRET during few intense gamma-ray bursts
suggested the idea that a large part of events could have
a high energy component, not observed so far due to the low fluxes
(Catelli 1997a).
Gamma-ray emission in the GeV-TeV energy range 
is predicted by some fireball models (see Baring 1997,
for a review). The study of the high energy part of the spectrum would be
of great importance to investigate the physical conditions of the emitting
region, restricting the range of fundamental parameters 
as the magnetic field, the density and the bulk Lorentz factor.

Unfortunately, due to the cosmological distances of the GRBs sources,
the high energy gamma-rays
would be absorbed by pair production on starlight photons during their
travel towards the Earth. 
According to Salomon and Stecker (1998), the flux of
gamma rays of energy $E>500$ GeV
is strongly reduced if the distance of the source is $z>0.1$.
Since the minimum redshift so far measured among 
six host galaxies is $z=0.695$ (Djorgovski 1999),
even assuming that GRBs emits gamma rays in the TeV region,
most of the spectra observed at Earth would cutoff at energies
less than few hundreds GeVs.

Observation of high energy GRBs can be
performed by ground based experiments, as air shower arrays, 
detecting the secondary particles produced in the atmosphere by primary
gamma-rays. 
Their large field of view ($\Omega \sim \pi$ sr)
and their duty cycle of almost 100$\%$ make them suitable
to observe unpredictable events as GRBs.
Due to the absorption of the high energy part of the spectrum, 
the detectors must be sensitive to gamma-rays of energy as low as 10-100
GeV. This can be achieved with two basic conditions:

{\it a)} by operating at very high mountain altitude, in order to increase
the number of detectable particles
(as an example,
the mean number of charged particles produced by a 100 GeV gamma-ray 
reaching the altitude of 2000 m is $n_c \sim 1.3$, while at 5000 m 
$n_c \sim 25$).

{\it b)} by disposing of a very large and "full-coverage" detection surface,
in order to detect the largest number of shower particles.

These conditions are fully satisfied by the ARGO-YBJ air shower detector.
In the following we present the sensitivity of ARGO-YBJ
to observe gamma-ray bursts as a function of the GRB spectral
characteristics.

\section{The ARGO-YBJ detector}

The ARGO-YBJ experiment has been conceived with the aim of detecting
small size atmospheric air showers. It is under  construction at the
Yangbajing High Altitude Cosmic Ray Laboratory (Tibet, China), at an
altitude of 4300 m above the sea level.
It consists of a central core made by a single layer of Resistive Plate
Chambers (RPCs) covering an area of $\sim$71 $\times$ 74 m$^2$,
sorrounded by an outer ring of 28 clusters of RPCs of area 42 m$^2$ each,
for a total sensitive area $A_d \sim$ 6100 m$^2$.
The detector is uniformely covered by a layer of lead 0.5 cm thick,
in order to increase the number of charged
particles by converting a fraction of the secondary photons,
and to reduce the time spread of the shower front.

One of the main field of research of ARGO-YBJ is 
gamma-ray astronomy in the energy range $E> 100$ GeV and  
gamma-ray bursts physics above 10 GeV.
A detailed description of the experiment, its capabilities and 
its physics goals are given by Abbrescia (1996)
and Bacci (1998).

\section{Detection of gamma-ray bursts}

The detection of high energy gamma-ray bursts can be
performed by the ARGO-YBJ detector by using two different modes of operation:

a) the "single particle" technique;

b) the "low multiplicity" technique.

In the following we discuss both methods and make a comparison
of their sensitivity in the energy range 10~GeV$<E<$1~TeV.

\subsection{The "single particle" technique (SP).}

An air shower array can be sensitive to primary energies as low as 10-100 GeV
operating in "single particle mode", i.e. recording all single 
secondary particles hitting the detector
with energy larger than the detection energy threshold $E_{th}$;
in this detection mode most of the events are due to solitary
muons and electrons 
of air showers generated by low energy cosmic rays.
A gamma-ray burst is detectable if the secondary particles
due to the gamma-rays interactions in the atmosphere give
a short time excess in the single particle counting rate, 
of amplitude larger than
the statistical fluctuations of the all-sky cosmic rays background.
The directions and energies of gamma-rays are not measurable;
however this technique could provide a measurement of the
total high energy flux and the temporal behaviour of the
high energy emission (Vernetto 1999, Aglietta 1999, Cabrera 1999).


The effective area to detect a primary gamma-ray
of energy $E$ and zenith angle $\theta$,
can be expressed as \mbox{$A_{eff}(E,\theta) \sim A_d  f_g n_e(E,\theta)$},
where $A_d$=6100 m$^2$ is the sensitive area,
$n_e(E,\theta)$ is the mean number of particles reaching the
detector level (with an energy larger than the detection threshold)
and $f_g$ is the gain factor due to the photons coversion in the lead layer
($f_g \sim 1.1$).
The curve A in Fig.1 shows the ARGO-YBJ effective area as a function of the
gamma-ray primary energy $E$, for a zenith angle $\theta$=20$^{\circ}$.

Given a GRB with an energy spectrum $dN_{\gamma}/dE$ (photons per unit 
area per unit energy) and zenith angle $\theta$ 
the number of events detected is
\mbox{$N_{SP} \ = A_d f_g cos \theta \int dN_{\gamma}/dE \ n_e(E,\theta) dE$}.

The signal is observable if the number of detected particles 
$N_{SP}$ is significantly 
larger than the background statistical fluctuations 
\mbox{$N_b=\sqrt{A_d \ B \ \Delta t}$},
where $B$ is background rate (events per unit area and unit time)
and $\Delta t$ is the GRB duration.
The measured single particle 
background rate at the Yangbajing site is $B \sim 1500$ events 
m$^{-2}$ s$^{-1}$. Requiring for the GRB signal
a minimum statistical significance of 4 standard deviations, 
the number $N_{SP}$ of events in ARGO-YBJ 
from a GRB of time duration $\Delta t$ = 1 s must be larger than
$\sim$1.2 10$^4$.

\subsection{The low multiplicity technique (LM).}

The low multiplicity technique (LM) consists in the detection 
of very small air showers, by requiring at least 6 fired pads
per shower (a {\it pad} is a detection unit of 56 $\times$ 56 cm$^2$,
see Abbrescia 1996).
The effective area $A_{eff}$ of ARGO-YBJ to detect primary gamma-rays and protons
using this technique has been obtained by simulations
and it is shown in Fig.1 for primaries with zenith angle 20$^{\circ}$
 (curve B for gamma-rays and curve C for protons).
The LM effective area for gamma-rays is 2-3 orders of magnitude
smaller than the "single particle" one (curve A), due to the 
higher number of particles required to satisfy the trigger condition
($\ge$6 instead of 1).
However the possibility to measure the primary arrival directions 
using the standard reconstruction technique of the shower front,
reduces significantly the background, limited to cosmic rays
with directions inside the angular error box.
The angular resolution for primaries with energy $E \sim$ 10 GeV
(defined as the opening angle around the source containing
the 70$\%$ of the signal showers) is $r \sim 5^{\circ}$
(Abbrescia 1996). Using the cosmic ray primary
proton spectrum measured by Honda (1995), the number of background
events with arrival directions in a cone of radius $r=5^{\circ}$ 
and zenith angle $\theta$=20$^{\circ}$ are expected to be 
$B_{LM} \sim$ 160 s$^{-1}$.
The number of events in ARGO-YBJ due to the burst is:
\mbox{$N_{LM} \ = 0.7 \int A_{eff} dN_{\gamma}/dE \ dE$}.
Requiring for the GRB signal
a minimum statistical significance of 4 standard deviations, 
the number $N_{LM}$ of events in ARGO-YBJ 
due to a GRB of time duration $\Delta t$ = 1 s and zenith angle 
$\theta$=20$^{\circ}$ must be larger than $\sim$50.

\subsection{Sensitivity to detect GRBs}

\begin{figure}[t]
\vfill \begin{minipage}{.49\linewidth}
\begin{center}
\mbox{\epsfig{file=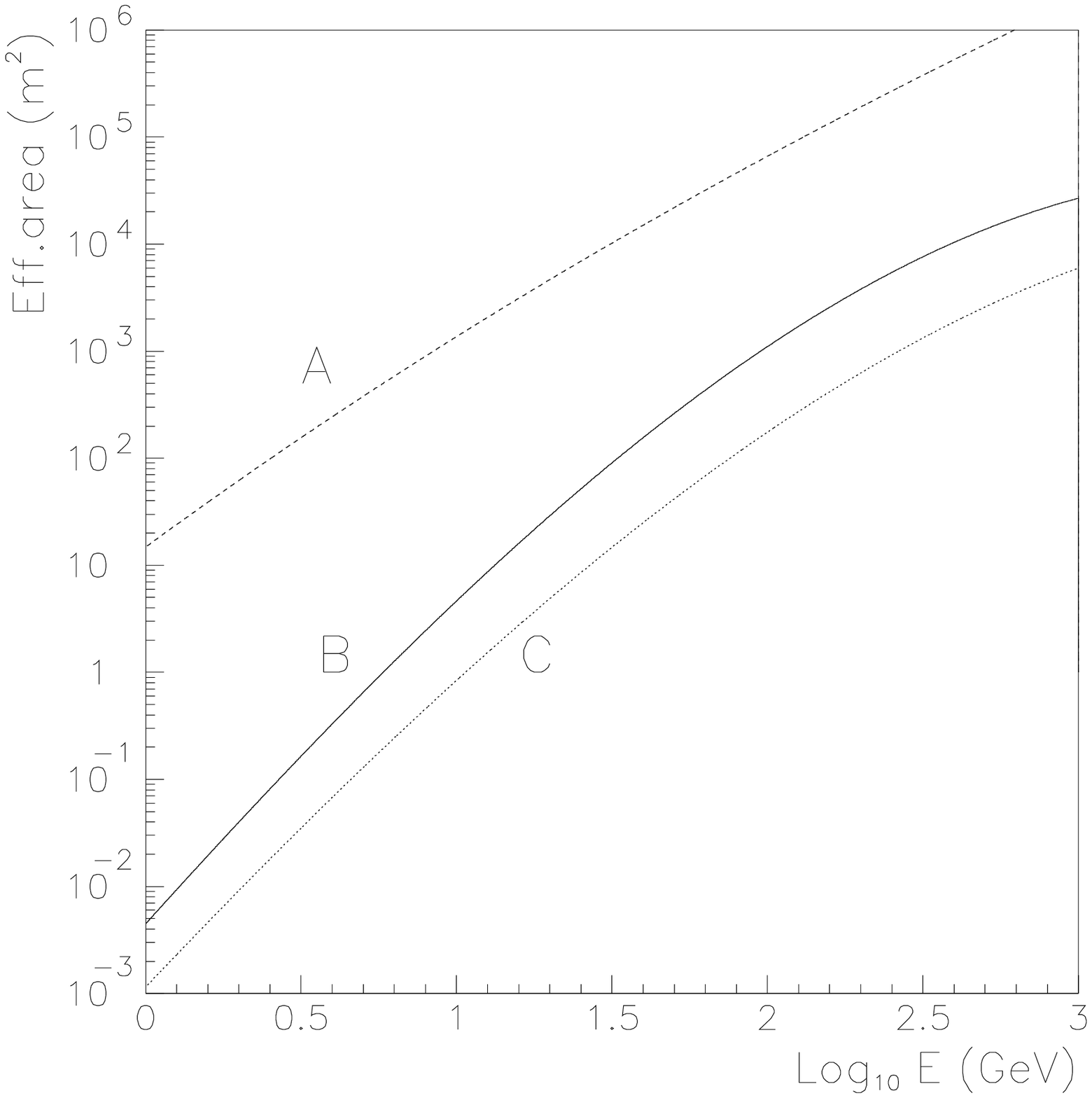,height=8.cm,width=8cm}}
\end{center}
\caption{Effective area of ARGO-YBJ to detect primary 
gamma-rays using the "single particle" technique
(curve A) and the "low multiplicity" technique (curve B); effective area to detect 
primary protons using the "low moltiplicity" technique (curve C).
The primary zenith angle is 20$^{\circ}$.}
\end{minipage}\hfill
\begin{minipage}{.49\linewidth}
\begin{center}
\mbox{\epsfig{file=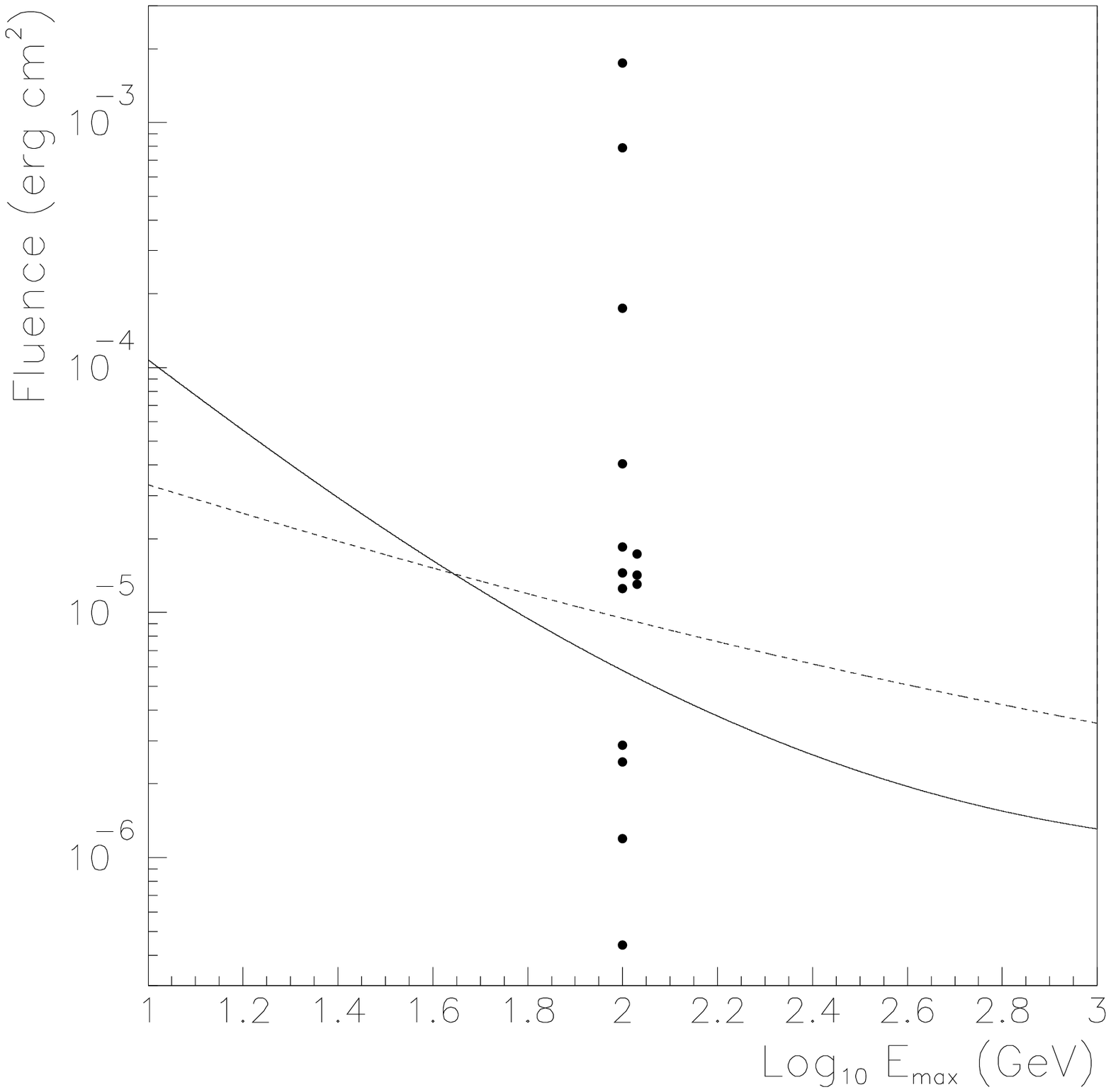,height=8.cm,width=8cm}}
\end{center}
\caption{Minimum energy fluence in the 1-100 GeV range observable by ARGO-YBJ,
as a function of the maximum energy of the specrum $E_{max}$, 
using the LM technique
(solid line) and the SP technique (dashed line). The points represent
 the extrapolations to 100 GeV of 14 EGRET spectra.}
\end{minipage}\vfill
\end{figure}

For simplicity we assume a burst giving a gamma-ray flux at the
top of the atmosphere as $dN/dE = K E^{-\alpha}$ photons m$^{-2}$
and a power law energy spectrum extending with unchanged slope 
up to a maximum energy $E_{max}$, with $E_{max}>$10 GeV.
This assumption is supported by
EGRET observations, which report power law spectra extending with
no visible cutoff up to the maximum energy determined by the
instrument sensitivity (in some cases above 1 GeV).
The average spectral slope observed in the 30~MeV-10~GeV region is $\alpha =
1.95 \pm 0.25$ (Dingus 1997).
Obviously a sharp cutoff at $E=E_{max}$ is unrealistic,
but for our purposes this simple parametrization can be adopted.
The energy cutoff can be due to an intrinsic cutoff at the source or/and
to the absorption of gamma-rays in the intergalactic space,
as previously mentioned. The latter effect could affect the spectra
at relatively low energy: according to Salomon and Stecker (1998),
gamma-rays of energy larger than $\sim$ 40 (100) GeV would be strongly absorbed
if the GRB distance is $z$=1.0 (0.5). 

In order to evaluate the ARGO-YBJ sensitivity,
it is convenient to work in terms of $F_{min}$,
defined as the minumum energy fluence in the energy range 1~GeV$\div E_{max}$ 
necessary to make a GRB observable by ARGO-YBJ, assuming that 
the spectrum extends with unchanged slope up to $E_{max}$.
Fig. 2 shows $F_{min}$ as a function of $E_{max}$, with $E_{max}$ in the range
10~GeV$\div$1~TeV, using the SP and the LM techniques.
The curves are given for a GRB duration $\Delta t$=1 s and 
a spectral slope: $\alpha$ = 2.0.
The minimum fluence for a different duration $\Delta t$ 
scales as $\sqrt{\Delta t}$.
The minimum required statistical significance of the signal
is $\sigma = 4$ standard deviations.

Obviously the sensitivity increases with $E_{max}$. The dependence
on $E_{max}$
is stronger for LM than for SP, given the different behaviour of 
the effective areas as a function of the gamma-ray energy
(moreover, the angular resolution improves with the energy,
an effect which is not accounted for in the present calculations).
This makes the SP technique more profitable 
when the energy spectrum has a relatively low energy cutoff,
in this case, for $E_{max}<$50~GeV. This value ranges between
35 and 70 GeV if the slope $\alpha$ varies from 1.5 to 2.5.

To compare the ARGO-YBJ sensitivity with the fluxes that can
be reasonably expected at
high energy, in the same figure we report the fluences in the
1$\div$100 GeV energy range obtained extrapolating (with the 
observed slopes)
the spectra measured by EGRET during the 15 events detected
by the TASC instrument (Catelli 1997b)(one of the events,
showing an unusal steep spectrum with $\alpha$=3.67, is not shown,
since the extrapolated fluence $F=10^{-10}$ erg cm$^{-2}$ falls well 
out of the plot). As can be seen in the figure, most of the events 
have an energy fluence larger than the ARGO-YBJ limits.

\section{Conclusions}

The ARGO-YBJ detector could observe GRBs in the energy range $E>10$ GeV
using the "single particle" technique (SP) and the "low multiplicity" technique
(LM). The SP method is more suitable for gamma-ray bursts with energy
spectra not extending more than $\sim$ 50 GeV, while the LM method
is preferable for more energetic spectra.
Adopting both techniques, ARGO-YBJ could detect GRBs with
energy fluence in the range 1$\div$100 GeV as low as 
$F\sim 10^{-6} \div 10^{-5}$ erg cm$^{-2}$,
if the spectral slope is $\alpha \sim$2.

\vspace{1ex}
\begin{center}
{\Large\bf References}
\end{center}
%
Abbrescia M. et al., Proposal of the ARGO experiment, 1996 

\ \ (can be downloaded
at the URL: http://www1.na.infn.it/wsubnucl/cosm/argo/argo.html)\\
Aglietta M. et al., 1999, Proc. Conf. GRBs in the afterglow era, in press\\
Bacci C. et al., Addendum to the ARGO Proposal, 1998 

\ \ (can be downloaded
at the URL: http://www1.na.infn.it/wsubnucl/cosm/argo/argo.html)\\
R.Cabrera et al., 1999, Proc. Conf. GRBs in the afterglow era, in press\\
Catelli J.R., Dingus B.L. and Schneid E.J., 1997, 25$^{th}$ ICRC Proc, 3, 33\\
Catelli J.R., Dingus B.L. and Schneid E.J., 1997, AIP Conf.Proc. 428, 309\\
Dingus B.L., Catelli J.R. and Schneid E.J., 1997, 25$^{th}$ ICRC Proc, 3, 30\\
Djorgovski S.G. et al 1999, GCN Circular 289 \\
Honda M et al., 1995, Phys.Rev.D 52, 4985\\
Salomon M.H. and Stecker F.W., 1998, ApJ 493, 547\\
Vernetto S., submitted to Astroparticle Phys, 1999, Astro-ph 9904324 \\
\end{document}